\documentclass[aps,prb,floatfix,epsfig,twocolumn,showpacs,preprintnumbers]{revtex4}
\usepackage{graphicx}
\usepackage{dcolumn}
\usepackage{bm}
\usepackage{amssymb}
\usepackage{amsmath}

\begin{document}

\title{Scaling Between Periodic Anderson and Kondo Lattice Models}
\author{R.~Dong$^{1}$, J.~Otsuki$^{2,3}$, and S.~Y.~Savrasov$^{1}$}
\affiliation{Department of Physics, University of California, Davis, California 95616, USA%
\\
$^{2}$Department of Physics, Tohoku University, Sendai 980-8578, Japan\\
$^{3}$Theoretical Physics III, Center for Electronic Correlations and
Magnetism, Institute of Physics, University of Augsburg, D-86135 Augsburg,
Germany}
\date{\today }

\begin{abstract}
Continuous--Time Quantum Monte Carlo (CT-QMC) method combined with Dynamical
Mean Field Theory (DMFT) is used to calculate both Periodic Anderson Model
(PAM) and Kondo Lattice Model (KLM). Different parameter sets of both models
are connected by the Schrieffer--Wolff transformation. For degeneracy $N=2$,
a special particle--hole symmetric case of PAM at half filling which always
fixes one electron per impurity site is compared with the results of the
KLM. We find a good mapping between PAM and KLM in the limit of large
on--site Hubbard interaction $U$ for different properties like self--energy,
quasiparticle residue and susceptibility. This allows us to extract
quasiparticle mass renormalizations for the f electrons directly from KLM.
The method is further applied to higher degenerate case and to realsitic
heavy fermion system CeRh$\text{In}_{5}$ in which the estimate of the
Sommerfeld coefficient is proven to be close to the experimental value.
\end{abstract}

\maketitle

\section{INTRODUCTION}

Computational study of heavy fermion materials\cite{Stewart} is a
challenging theoretical problem. These systems are a subset of intermetallic
compounds that have a low--temperature specific heat whose linear term is up
to 1000 times larger than the value expected from the free-electron theory.
The heavy fermion behavior has been found in rare earth and actinide metal
compounds at very low temperatures (typically less than 10 K) in a broad
variety of states including metallic, superconducting, insulating and
magnetic states\cite{Misra}.

The physics of the heavy fermion systems is controlled by the
antiferromagnetic interactions of local magnetic moments residing on the
rare earth or actinide atoms with the sea of conduction electrons. The
theoretical problem of a localized spin interacting with the conduction
electrons is the celebrated Kondo problem\cite%
{Kondo,Abrikosov,Suhl,Nagaoka,Willson,Hewson} whose solution is one of the
outstanding achievements of many-body physics. It describes how the local
spin is compensated as the temperature falls below a characteristic Kondo
temperature. Something similar occurs in the heavy fermion materials which
represents an array of such spins forming a Kondo lattice.

In this regime, each f orbital is occupied by a fixed number of electrons,
and all types of charge fluctuations are approximately frozen due to a large
Coulomb repulsion penalty that the system pays when the electron is
added/removed from the shell. Therefore the low--energy degrees of freedom
are provided by localized spins only and the corresponding model is known as
the Kondo Lattice Model (KLM)\cite%
{Lacroix,Coleman,Doniach,Caprara,Tsunetsugu,Lavagna,Si,Nolting,Senthil}. The
KLM effective Hamiltonian is obtained by using a second--order perturbation
with respect to hybridization\cite{SchriefferWolf} of a more general
Periodic Anderson Model (PAM) \cite{Jullian,Rice,Scalapino}where the
localized f--electrons can exchange with the conduction electrons bath thus
allowing both charge and spin fluctuations to occur. The introduction of the
limit of infinite dimensions and subsequent development of the dynamical
mean field theory (DMFT) has allowed to study the properties of both models
in a systematic manner\cite{Vollhard,Gabi,Mark,RMP,Jarrell,Thomas,Matsumoto}

Due to developments in the electronic structure theory for strongly
correlated systems based on a combination of Density Functional Theory (DFT)
in its local density approximation (LDA)\ and DMFT\cite{RMP2}, studies of
real heavy fermion materials have recently appeared in the literature\cite%
{Held,Haule,Shim}. Here the development of Continuous Time Quantum Monte
Carlo Method (CT--QMC) for solving corresponding Anderson Impurity problem
has played a central role \cite{Rubtsov,Werner,HauleQMC,Gull}. These
calculations are extremely computationally demanding especially for the f
elements such as Plutonium\cite{PuAm,HaulePu} where a large number of atomic
states needs to be kept in the calculation.

We have recently proposed a simplified approach\cite{MatsumotoPRL} where
instead of full solution of the Anderson impurity model, a corresponding
Kondo Impurity (or more general Coqblin--Schrieffer Impurity\cite{Coqblin})
is studied to explore low--energy physics of heavy fermion materials\cite%
{MatsumotoPRB1,MatsumotoPRB2} using most recently developed CT--QMC
algorithm for this problem\cite{Otsuki1,Otsuki2,Otsuki3}. In this regard, an
interesting question arises on how exactly the scaling between Anderson and
Kondo impurity models occurs and whether the low--energy properties of heavy
fermion systems such as electronic mass enhancement and associated with it
linear specific heat coefficient can be recovered from a restricted solution
provided by the conduction electron self--energies available within the KLM.
Such scaling behavior has been explored \cite{Wilkins} for the temperature
dependent susceptibility of the symmetric Anderson model using numerical
renormalization group techniques\cite{Willson}, where a precise mapping has
been found to spin--$\frac{1}{2}$ Kondo Hamiltonian. Here we explore a
similar mapping between single--particle functions such as the self--energy
where upon increasing the value of on--site Coulomb repulsion $U$, we report
a convergence of the conduction electron self--energy extracted from the
solution of the PAM to the one obtained within KLM. We use CT--QMC method
for the corresponding impurity models and the Dynamical Mean Field Theory
for achieving self--consistent solution of the lattice problem. We utilize
an inverse relationship to extract the f--electron self--energies and
monitor how the low--frequency behavior of the AIM converges to the KLM\
limit. Our obtained mapping allows the extraction of the mass
renormalization of heavy quasiparticles directly from the solution of the
Kondo lattice Hamiltonian.

As an illustration, we consider an electronic structure of CeRhIn$_{5}$
where we compute hybridization functions of the f--electrons with conduction
bath and evaluate Kondo exchange coupling. We subsequently solve the Kondo
Lattice Model with CT--QMC and DMFT,\ compute conduction electron
self--energies, and then use the inverse mapping obtained from our analysis
of the model Hamiltonian to evaluate electronic mass enhancement and
specific heat coefficient of this system. Our theoretical results are
compared with available experimental data.

This article is organized as follows. In Section II, we discuss the mapping
between periodic Anderson model and Kondo lattice model in the limit of
large $U,$ and provide the results for electronic self--energies,
quasiparticle residues and susceptibilities. In Section III, application is
presented to evaluate electronic mass enhancement and Sommerfeld's
coefficient for CeRhIn$_{5}$. Section IV is the conclusion.

\section{MODEL CALCULATION}

\subsection{Periodic Anderson and Kondo Lattice Models}

One of the popular models to describe the physics of heavy fermion materials
is a periodic Anderson model\cite{Jullian,Rice,Scalapino}. The effective
Hamiltonian is given by%
\begin{equation}
\begin{split}
H_{\text{PAM}}=& \sum_{\mathbf{k}\sigma }\epsilon _{\mathbf{k}\sigma }c_{%
\mathbf{k}\sigma }^{\dagger }c_{\mathbf{k}\sigma }+\epsilon
_{f}\sum_{i\sigma }f_{i\sigma }^{\dagger }f_{i\sigma }+U\sum_{i}n_{i\uparrow
}^{f}n_{i\downarrow }^{f} \\
& +\sum_{i\mathbf{k}\sigma }V_{\mathbf{k}}(c_{\mathbf{k}\sigma }^{\dagger
}f_{i\sigma }+H.c.)
\end{split}%
\end{equation}%
where $c_{\mathbf{k}\sigma }^{\dagger }$ ($c_{\mathbf{k}\sigma }$) creates
(destructs) a conduction electron with momentum $\mathbf{k}$, spin (and
orbital) $\sigma $ and dispersion $\epsilon _{\mathbf{k}\sigma }$; $%
f_{i\sigma }^{\dagger }$ ($f_{i\sigma }$) creates (destructs) an $f$
electron with spin $\sigma $ and energy $\epsilon _{f}$ on site $i$; $%
n_{i\uparrow }^{f}$ ($n_{i\downarrow }^{f}$) is the number operator for $f$
electron at lattice site $i$ with spin up (down); $U$ is the on--site
Coulomb repulsion; $V_{\mathbf{k}}$ is the hybridization between $f$
electrons and conduction electrons which we assume to be $\mathbf{k}$%
--independent, $V_{\mathbf{k}}=V,$ for simplicity.

In systems where the Hubbard $U$ is large, the charge fluctuations become
effectively frozen and the ground state wave function has a little weight of
configurations with the number of f--electrons different from its average
number $\bar{n}_{f}$. This results in transforming the PAM\ Hamiltonian
which eliminates the hybridization term in first order by Schrieffer--Wolff
transformation. The second--order in $V$ Hamiltonian is a famous Kondo
Lattice Hamiltonian\cite%
{Lacroix,Coleman,Doniach,Caprara,Tsunetsugu,Lavagna,Si,Nolting,Senthil} that
describes interaction between spins of localized and conduction electrons 
\begin{equation}
H_{\text{KLM}}=\sum_{\mathbf{k}\sigma }\epsilon _{\mathbf{k}\sigma }c_{%
\mathbf{k}\sigma }^{\dagger }c_{\mathbf{k}\sigma }+J_{K}\sum_{i}S_{i}\cdot
\sigma
\end{equation}%
where $S_{i}$ represents the localized spin of the f electron at the $i$
site, the $\sigma $ is the spin operator of the itinerant conduction
electron and $J_{K}$ is the Kondo coupling constant.%
\begin{equation}
J_{K}=V^{2}(\frac{1}{-\epsilon _{f}}+\frac{1}{\epsilon _{f}+U}).  \label{SW}
\end{equation}%
For the half--filled case, $\epsilon _{f}=-U/2$, and $J_{K}$ is simplified
to $4V^{2}/U$.

\subsection{Dynamical Mean Field Theory}

Solutions of both models in a general case represent a complicated numerical
problem. Using dynamical mean field theory, the algorithm breaks down into
(i) the solution of the corresponding (Anderson or Kondo) impurity problem
and (ii) the self--consistency loop over hybridization functions which
enforces lattice periodicity\cite{RMP}.

For the periodic Anderson model, the DMFT evaluates the local Green function
for heavy electrons%
\begin{equation*}
G_{f}^{(loc)}(i\omega _{n})=\sum_{\mathbf{k}}\left[ i\omega _{n}-\epsilon
_{f}-\Sigma _{f}(i\omega _{n})-\frac{V^{2}}{i\omega _{n}-\epsilon _{\mathbf{k%
}}}\right] ^{-1}.
\end{equation*}%
Then, the bath Green function is defined%
\begin{equation*}
G_{f}^{(0)-1}(i\omega _{n})=G_{f}^{-1}(i\omega _{n})+\Sigma _{f}(i\omega
_{n})\equiv i\omega _{n}-\epsilon _{f}-\Delta (i\omega _{n})
\end{equation*}%
and used as an input to the impurity solver. The latter produces an impurity
Green function 
\begin{equation*}
G_{f}^{(imp)}(i\omega _{n})=\frac{1}{i\omega _{n}-\epsilon _{f}-\Delta
(i\omega _{n})-\Sigma _{f}(i\omega _{n})}
\end{equation*}%
from where, a new self--energy can be found%
\begin{equation*}
\Sigma _{f}(i\omega _{n})=G_{0}^{-1}(i\omega _{n})-G_{f}^{(imp)-1}(i\omega
_{n}).
\end{equation*}%
The process is repeated by recalculating the lattice Green function with the
new self--energy. The self--consistency condition is when 
\begin{equation*}
G_{f}^{(imp)}(i\omega _{n})=G_{f}^{(loc)}(i\omega _{n}).
\end{equation*}

The Kondo lattice Hamiltonian can be obtained by considering the limit $%
V^{2}\rightarrow \infty ,U\rightarrow \infty ,\epsilon _{f}\rightarrow
-\infty $ while keeping $V^{2}/\epsilon _{f}=const.$ First, define a local
Green function for conduction electrons

\begin{equation*}
G_{c}^{(loc)}(i\omega _{n})=\sum_{\mathbf{k}}\left[ i\omega _{n}-\epsilon _{%
\mathbf{k}}-\Sigma _{c}(i\omega _{n})\right] ^{-1}
\end{equation*}%
where conduction electron self energy is given by

\begin{equation*}
\Sigma _{c}(i\omega _{n})=\frac{V^{2}}{i\omega _{n}-\epsilon _{f}-\Sigma
_{f}(i\omega _{n})}.
\end{equation*}%
Thus, one can iterate over conduction electron quantities: The bath Green
function

\begin{equation*}
G_{c}^{(0)-1}(i\omega _{n})=G_{c}^{(loc)-1}(i\omega _{n})+\Sigma
_{c}(i\omega _{n})
\end{equation*}%
serves as the input to the Kondo impurity solver. The latter produces an
impurity Green function $G_{c}^{(imp)}(i\omega _{n})$ from which the new
conduction electron self--energy is found. The process is repeated by
reevaluating $G_{c}^{(loc)}(i\omega _{n}).$The self--consistency condition
is when 
\begin{equation*}
G_{c}^{(imp)}(i\omega _{n})=G_{c}^{(loc)}(i\omega _{n}).
\end{equation*}

Several powerful methods such as exact diagonalization or numerical
renormalization group techniques have been developed in the past to deal
with the impurity models. In this work we utilize a Continuous Time Quantum
Monte Carlo method\cite{Rubtsov,Werner,HauleQMC,Gull} that was originally
proposed to deal with Anderson impurities but has been recently generalized
for Kondo (Coqblin--Schrieffer) type of impurities \cite{Otsuki1}.

The density of states of conduction electrons is an input to the simulation.
Despite realistic materials may have complex band structures, we use a
simple constant density of states to gain the physical insight from these
calculations. The half--bandwidth $D$ is set to 1 which provides the
corresponding units. As we are looking for a mapping between the two models
in the regime of large $U$, we first fix the Kondo coupling $J_{K}$ to some
predetermined value. There are typically two phases that emerge in the KLM:
the antiferromagnetic RKKY\ phase and the paramagnetic Fermi liquid phase,
which compete with each other on the scale of $J_{K}$ \cite{Doniach} . We
are mainly interested in the Fermi liquid behavior and consider the value of 
$J_{K}=0.3$ in all our calculations. Second, we study two cases with the
effective f--electron degeneracies $N=2$ and $N=4$. For the case $N=2$, the
only non--trivial occupancy of the f--orbital is 1 which for the
particle--hole symmetric placement of the conduction electron band results
in the condition $\epsilon _{f}=-U/2$, for the $f$ orbital to be
half--filled. Although the system becomes a band (Kondo) insulator in this
case, the f--electrons states are strongly renormalzed by correlations which
is the basis for the comparison of these two models.

\subsection{DMFT\ Solutions for $N=2$}

\subsubsection{Self-energies}

\begin{figure}[tbp]
\includegraphics[width=3.5in]{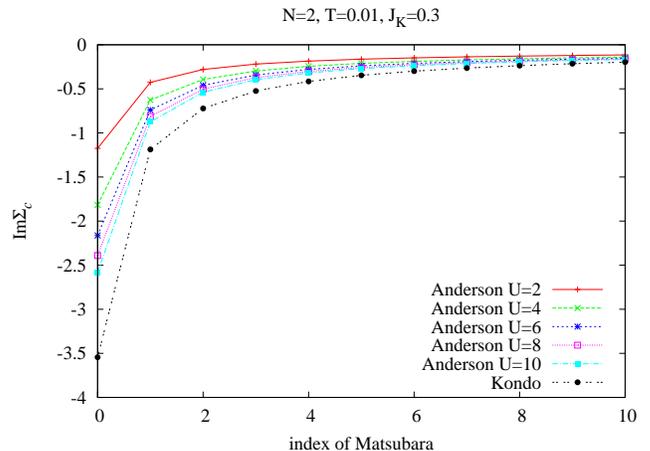}
\caption{(color online) Conduction electron self--energy of the periodic
Anderson model with $N=2$ calculated for several values of $U$ and the
conduction electron self--energy of the Kondo lattice model that
correspoinds to $U\rightarrow \infty $ limit.}
\label{fig-ImC}
\end{figure}

We first obtain the f--electron self--energy from the PAM calculation. Then
we extract the conduction electron self--energy and compare with the data of
the KLM simulation. We present such comparison in Fig. \ref{fig-ImC} where
we plot $\Im \Sigma _{c}(i\omega _{n})$ for several values of $U$ and $%
U\rightarrow \infty $ limit corresponding to KLM. We monitor a slow
convergence of the PAM\ self--energy towards its KLM value, although even
for $U=10$ the discrepancy between the two is still noticeable.

We now address the question of $U\rightarrow \infty $ limit for the
f--electron self--energies$,$ which numerically corresponds to the Kondo
regime, and compare these data with our scaling behavior established
analytically. First, notice that the low--frequency expansion for both
f--electron and conduction electron quantities can be derived without a
problem 
\begin{equation*}
\Sigma _{f,c}(i\omega _{n})=\Sigma _{f,c}(0)+i\omega _{n}(1-z_{f,c}^{-1})
\end{equation*}%
which leads us to%
\begin{eqnarray}
\Sigma _{f}(0) &=&-\epsilon _{f}-\frac{V^{2}}{\Sigma _{c}(0)}  \notag \\
z_{f} &=&\frac{[\mathfrak{R}\Sigma _{c}(0)]^{2}}{V^{2}}\frac{z_{c}}{1-z_{c}}.
\label{zf2zc}
\end{eqnarray}%
This formula has been used in our recent LDA+DMFT\ work to extract mass
renormalization parameters in several heavy fermion compounds \cite%
{MatsumotoPRB2} using simulations with the Kondo lattice.

Unfortunately, this approach will not work for the model considered here
since for the particle--hole symmetric case of the Kondo insulator, the
conduction electron self--energies diverge to produce an energy gap in the
excitational spectrum. We therefore look for a scaling behavior in a
different way. We write 
\begin{equation}
\Sigma _{f}(i\omega _{n})=i\omega _{n}-\epsilon _{f}-\frac{V^{2}}{\Sigma
_{c}(i\omega _{n})}  \label{ctof}
\end{equation}%
and noticing that we target the $U\rightarrow \infty $ limit, we replace $%
V^{2}$ with $\frac{1}{4}J_{K}U$, and divide both parts by $U$. For the
imaginary part we obtain the following scaling behavior 
\begin{equation}
\frac{\Im \Sigma _{f}(i\omega _{n})}{U}=\frac{J_{K}\Im \Sigma _{c}(i\omega
_{n})}{4|\Sigma _{c}(i\omega _{n})|^{2}}  \label{simple}
\end{equation}%
that expresses the large $U$ limit of the PAM\ self--energy via the
quantities available within KLM.

\begin{figure}[tbp]
\includegraphics[width=3.5in]{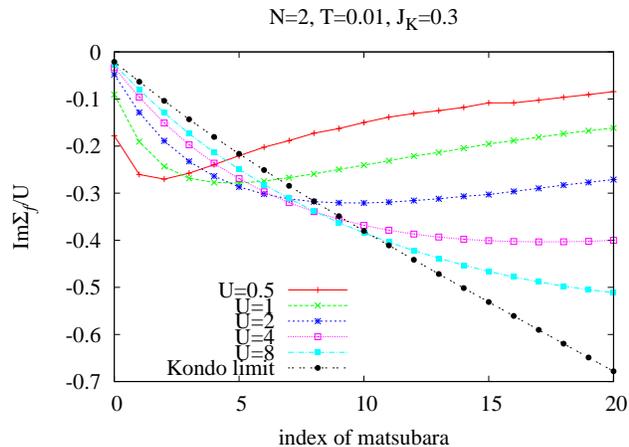}
\caption{(color online) The convergence for the f--electron self-energy
obtained from the periodic Anderson model with $N=2$ upon increase in the
interaction $U$. The limiting behavior of this quantity extracted from the
solution of the Kondo lattice model is also plotted. }
\label{fig-ImefU}
\end{figure}

The self--energies from both models can now be directly compared. Fig. \ref%
{fig-ImefU} shows the behavior of $\Im \Sigma _{f}(i\omega _{n})/U$ for
several values of $U$ together with the corresponding data extracted from
KLM. From the figure we see that as $U$ increases the PAM\ self--energy
converges to that of the KLM. The plot actually includes both the
intermediate regime and the Kondo regime. When $U<2$ the two Hubbard bands
are within the conduction electron band which has a bandwidth of 2 in our
units. In this case the self--energy deviates more than the data for $%
U\geqslant 2$. As $U$ goes to 8, the self--energies from both models
collapse at low energies.

Finally, we notice here that the $U\rightarrow \infty $ limit produces $\Im
\Sigma _{f}(i\omega _{n})/U$ that grows linearly with the frequency which is
exactly the case of the self--energies obtained in slave--boson type of
methods for solving the impurity problems\cite{SB,SB2} or within
quasiparticle Gutzwiller approximation \cite{Gutz}.

\subsubsection{Quasiparticle Residues}

Our self--energy results show qualitative convergence between the two
models. To get quantitative agreement we further check the quasiparticle
residues that we extract from the low frequency behavior of the
self--energies. These are related to renormalized effective masses for the
quasiparticles responsible for the enhanced specific heat coefficient which
is one of the central properties of systems with heavy fermions.

The CT--QMC algorithm works on an imaginary time axis, which after the
Fourier transformation gives us the data on the imaginary frequency axis.
Despite problems associated with numerical noise that prevents us to extract
accurate data at real frequencies using analytical continuation, we can find
the quasiparticle residue from the imaginary axis data as follows:

\begin{equation}
z=(1-\frac{\partial \Im \Sigma (i\omega _{n})}{\partial (i\omega _{n})}%
|_{\omega _{n}\rightarrow 0})^{-1}.  \label{z}
\end{equation}%
For the KLM, according to Eq. (\ref{ctof}), the $z_{f}$ can be written as:

\begin{equation}
z_{f}^{KLM}=-\frac{4{\pi }T}{J_{K}U}\frac{|\Sigma _{c}({i\pi }T)|^{2}}{\Im
\Sigma _{c}(i{\pi }T)}.  \label{zklm}
\end{equation}%
Although this expression is actually valid only for $U=\infty $, where $%
z_{f}^{KLM}$ becomes zero, we expect that it gives an approximate value for
PAM with $U<\infty .$

\begin{figure}[tbp]
\includegraphics[width=3.5in]{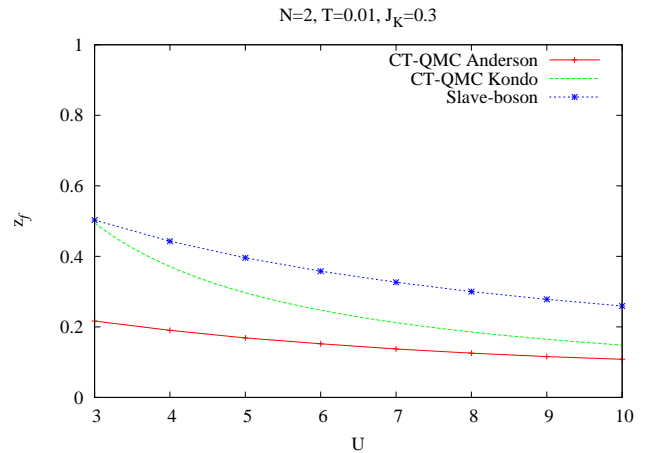}
\caption{(color online) Comparison between quasiparticle residues $z_{f}$
calculated as a function of Hubbard $U$ using periodic Anderson model with $%
N=2$, Kondo lattice model as well as a slave--boson method described in Ref. 
\onlinecite{SB2}.}
\label{fig-zf}
\end{figure}

We present our comparisons between the two quantities in Fig. \ref{fig-zf}
where we plot the quasiparticle residue extracted from PAM as a function of $%
U$ as well as the one extracted from the KLM according to Eq. (\ref{zklm}).
Also plotted for comparison is the quasiparticle residue calculated using
the slave--boson method as described in Ref. \onlinecite{SB2}.

From the plots, we can quantitatively see the convergence from\ the result
obtained from PAM to the one obtained by KLM. When $U$ reaches $10$, the KLM
overestimates $z_{f}$ of the PAM data by about 30\%. While the slave--boson
method, a very fast calculation, demonstrates a similar behavior, it
overestimates the PAM data by about 100\%.

\subsubsection{Susceptibilities}

Since the KLM freezes the spatial fluctuations but keeps the essential
magnetic properties, the corresponding spin susceptibilities $\chi (T)$
should agree between the PAM\ and KLM at least for low temperatures. Indeed,
these quantities have been calculated and compared against each other for
the corresponding impurity models using numerical renormalization group
methods long time ago \cite{Wilkins} where a precise mapping between
Anderson impurity and spin--$\frac{1}{2}$ Kondo impurity has been observed. 
\begin{figure}[tbp]
\includegraphics[width=3.5in]{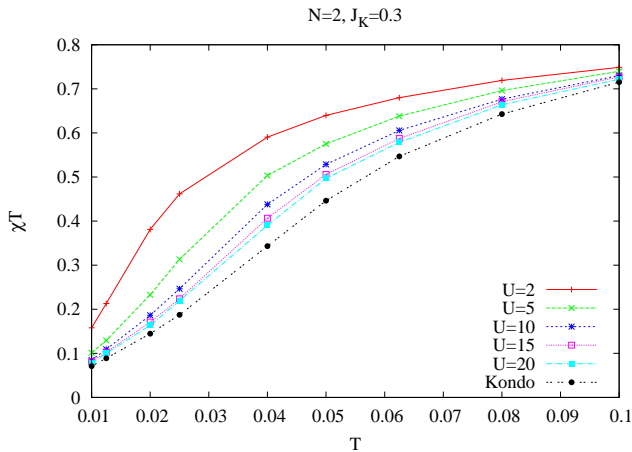}
\caption{(color online) Calculated using the periodic Anderson model with $%
N=2$ temperature dependence of spin susceptibility (times the temperature)
for several values of Hubbard $U$ as well as the data extracted for the
Kondo lattice model.}
\label{fig-chi}
\end{figure}

We present our own calculations in Fig. \ref{fig-chi} where we plot $T\chi
(T)$ against the temperature for several values of U calculated using the
PAM as well as the data extracted from the KLM. The comparison shows a nice
convergence for susceptibility within our chosen temperature range. The
deviation may result from the combination of thermal effect and charge
fluctuations. We see that the convergence is worse here than that for the
quasiparticle residues discussed earlier, but we believe that the
susceptibilities should map precisely to each other at lower temperatures.

\subsection{DMFT\ Solutions for $N=4$}

The $N=2$ case with one localized electron leads to the Kondo insulator
state which is topologically special. However, our method can be generalized
to larger orbital degeneracy, where any integer occupancy of the f--shell
can be explored. Also the Coqblin--Schrieffer model is more favorable. Below
we consider the case with $N=4$ and $n_{f}=1$ which is away from
particle--hole symmetry. As is the case with $N=2$, we fix the value of the
Kondo coupling $J_{K}$ to 0.3. For each value of $U$ that we input to the
PAM calculation, there are two remaining parameters, the impurity level $%
\epsilon _{f}$, and the value of hybridization $V^{2}$ that should be
searched for to obtain $n_{f}=1,$ $J_{K}=0.3$.

Conduction electron self--energies $\Im \Sigma _{c}(i\omega _{n})$
calculated within PAM for several values of $U$ as well as within KLM
corresponding to $U\longrightarrow \infty $ limit are compared in Fig.\ref%
{N_4_Sigma_c}. We see that the convergency of $\Im \Sigma _{c}(i\omega _{n})$
is rather slow\ when $U$ increases similar to the $N=2$ case presented in
Fig. \ref{fig-ImC}. The low frequency behavior of $\Im \Sigma _{c}(i\omega
_{n})$ shows that the hybridization gap is no longer opened at the Fermi
energy and the system remains metallic contrary to the particle--hole
symmetric case of the Kondo insulator where $\Im \Sigma _{c}(i\omega _{n})$
diverges at $i\omega _{n}\rightarrow 0$ as seen on Fig. \ref{fig-ImC}. Here,
the low--frequency slopes determine quasiparticle residues $z_{c}$ for
conduction electrons which display a somewhat faster convergence to the
Kondo limit upon increase in $U$.

\begin{figure}[tbp]
\includegraphics[width=3.5in]{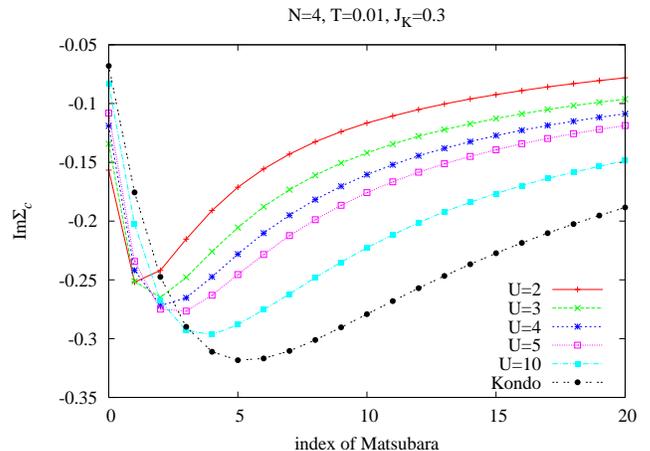}
\caption{(color online) Conduction electron self--energy of the periodic
Anderson model with $N=4$ calculated for several values of $U$ and the
conduction electron self--energy of the Kondo lattice model that
correspoinds to $U\rightarrow \infty $ limit.}
\label{N_4_Sigma_c}
\end{figure}

To compare how $\Sigma _{f}(i\omega _{n})$ scales to the Kondo limit, we
start from Eq. (\ref{ctof}) , take its imaginary part and divide by $U$ on
both sides. Using Eq. (\ref{SW}), the formula becomes:

\begin{equation}
\frac{\Im \Sigma _{f}(i\omega _{n})}{U}=\frac{i\omega _{n}}{U}+J_{K}\eta
(1-\eta )\frac{\Im \Sigma _{c}(i\omega _{n})}{|\Sigma _{c}(i\omega _{n})|^{2}%
}  \label{general}
\end{equation}%
where $\eta =-\frac{\epsilon _{f}}{U}$ is a dimensionless parameter which
affects the electron counting at impurity site.

\begin{figure}[tbp]
\includegraphics[width=3.5in]{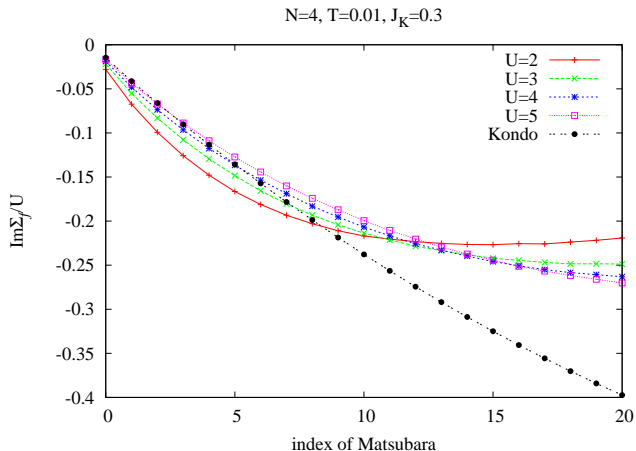}
\caption{(color online) The convergence for the f--electron self-energy
obtained from the periodic Anderson model with $N=4$ upon increase in the
interaction $U$. The limiting behavior of this quantity extracted from the
solution of the Kondo lattice model is also plotted. }
\label{N_4_Sigma_f}
\end{figure}

Fig. (\ref{N_4_Sigma_f}) presents the behavior of $\Im \Sigma _{f}(i\omega
_{n})/U$ calculated within PAM for several values of $U$ together with the
corresponding data extracted from KLM. From the figure we see that as $U$
increases, the PAM\ self--energy maps into its $U\longrightarrow \infty $
limit of the KLM.

\begin{figure}[tbp]
\includegraphics[width=3.5in]{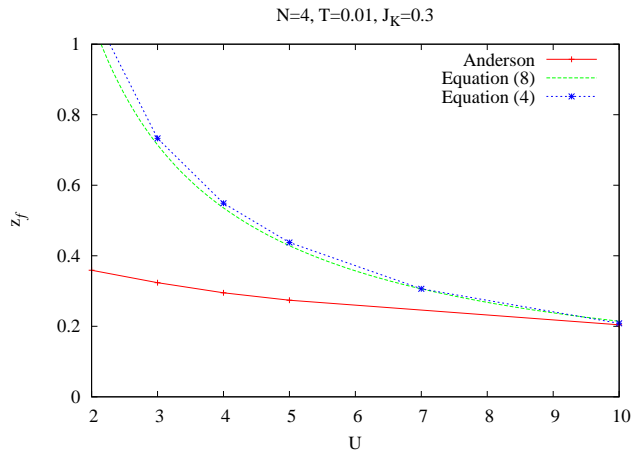}
\caption{(color online) Comparison between quasiparticle residues $z_{f}$
calculated as a function of Hubbard $U$ using periodic Anderson model with $%
N=4$, and the values extracted from the Kondo lattice model using two
different approaches described in text.}
\label{N_4_Z_f}
\end{figure}

We have finally extracted the values for quasiparticle residues from the
low--frequency slopes of $\Im \Sigma _{f}(i\omega _{n})$ which can be
compared with the values of $z_{f}$ that we obtain from the KLM calculation
either using the approach that leads us to Eq.(\ref{zklm}) or using the
low--frequency behavior of $\Im \Sigma _{c}(i\omega _{n})$ that leads us to
Eq. (\ref{zf2zc}). We present such comparison in Fig. \ref{N_4_Z_f} where
the behavior of $z_{f}$ is plotted against Hubbard $U.$ We see that starting
from $U=10$, the quasiparticle residues computed from PAM and KLM become
very close to each other.

\section{APPLICATION TO\ CeRhIn$_{5}$}

Realistic heavy fermion materials have much more complicated electronic
structures than we used in our model calculations. The f--orbitals are
14--fold degenerate and split in the presence of spin--orbit coupling and
crystal--field effects. This makes LDA+DMFT calculations with full solution
of Anderson impurity problem dramatically heavy. With the advantage of
reaching lower temperature range, LDA+DMFT\ simulations with Kondo (or
Coqblin--Schrieffer) impurity was established \cite{MatsumotoPRL}. In this
method, the first step is to find the a local hybridization function $\Delta
_{\alpha }(\epsilon )$ between fully localized $f$ and conduction $spd$
electrons, where subscript $\alpha $ refers to a particular representation
that tries to diagonalize a general matrix $\Delta _{m^{\prime }\sigma
^{\prime }m\sigma }(\epsilon )$ by taking advantage of spin--orbit and
crystal--field symmetries. This is achieved by using the so called Hubbard
I\ approximation\cite{Hubbard1} where purely atomic f--electron self--energy
is inputted to the LDA+DMFT calculation. Knowing $\Delta _{\alpha }(\epsilon
)$, the Kondo coupling constant $J_{K}$ and the initial Green's function of
conduction electrons can be extracted.

\begin{equation}
G_{\alpha }(i\omega _{n})=\int_{-D_{\text{cutoff}}}^{D_{\text{cutoff}%
}}d\epsilon \frac{\Im \Delta _{\alpha }(\epsilon )}{i\omega _{n}-\epsilon }%
/\int_{-D_{\text{cutoff}}}^{D_{\text{cutoff}}}d\epsilon \Im \Delta _{\alpha
}(\epsilon ),
\end{equation}%
and 
\begin{equation}
V_{\alpha }^{2}=\frac{1}{\pi }\int_{-D_{\text{cutoff}}}^{D_{\text{cutoff}%
}}d\epsilon \frac{\Im \Delta _{\alpha }(\epsilon )}{N_{F}}.
\end{equation}%
These are required for solving realistic Kondo lattice problem with $J_{K}$
calculated from the Schrieffer--Wolff transformation, Eq. (\ref{SW}).

Our target material is $\text{CeRhIn}_{5}$ which is believed to have most
localized f--electrons in the 115 family. The spin--orbit coupling and
crystal fields of the tetragonal structure effectively reduce the degeneracy
and make the $\Gamma _{7}$ doublet of the $j=5/2$ state to be the ground
state. Therefore at very low temperatures, a single localized f--electron
resides at the $\Gamma _{7}$ doublet and we have the $N=2$ case discussed
above in the model calculation.

\begin{table}[tb]
\caption{Calculated LDA density of states at the Fermi energy, $N(0)$
(states/eV/cell), estamted quasiparticle residue $z_{f}$ for the
f--electrons as well as predicted and experimental values of the Sommerfeld
coefficient $\protect\gamma $ (mJ/mol/$\text{K}^{2}$) for CeRhIn$_{5}.$}
\label{CeRhIn5}\centering%
\begin{tabular}{ccccc}
\hline\hline
materials & $N(0)_{\text{LDA}}$ & $z_{f}$ & $\gamma $ & $\gamma _{exp}$ \\ 
\hline
$\text{CeRhIn}_{5}$ & 2.21 & 0.01255 & 414 & 400$^a$ \\ \hline\hline
\end{tabular}%
\newline
$^a$Ref.\onlinecite{Hegger}
\end{table}

\begin{figure}[tbp]
\includegraphics[width=3.5in]{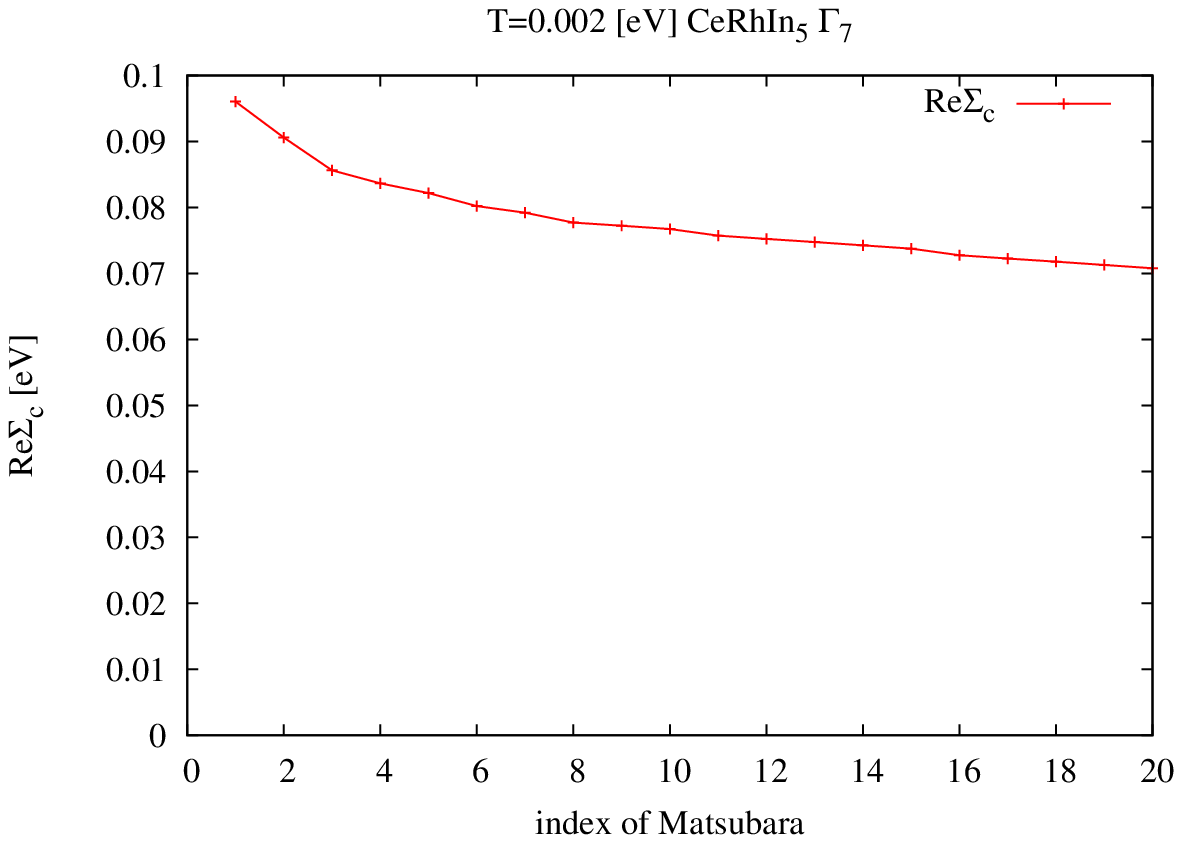} %
\includegraphics[width=3.5in]{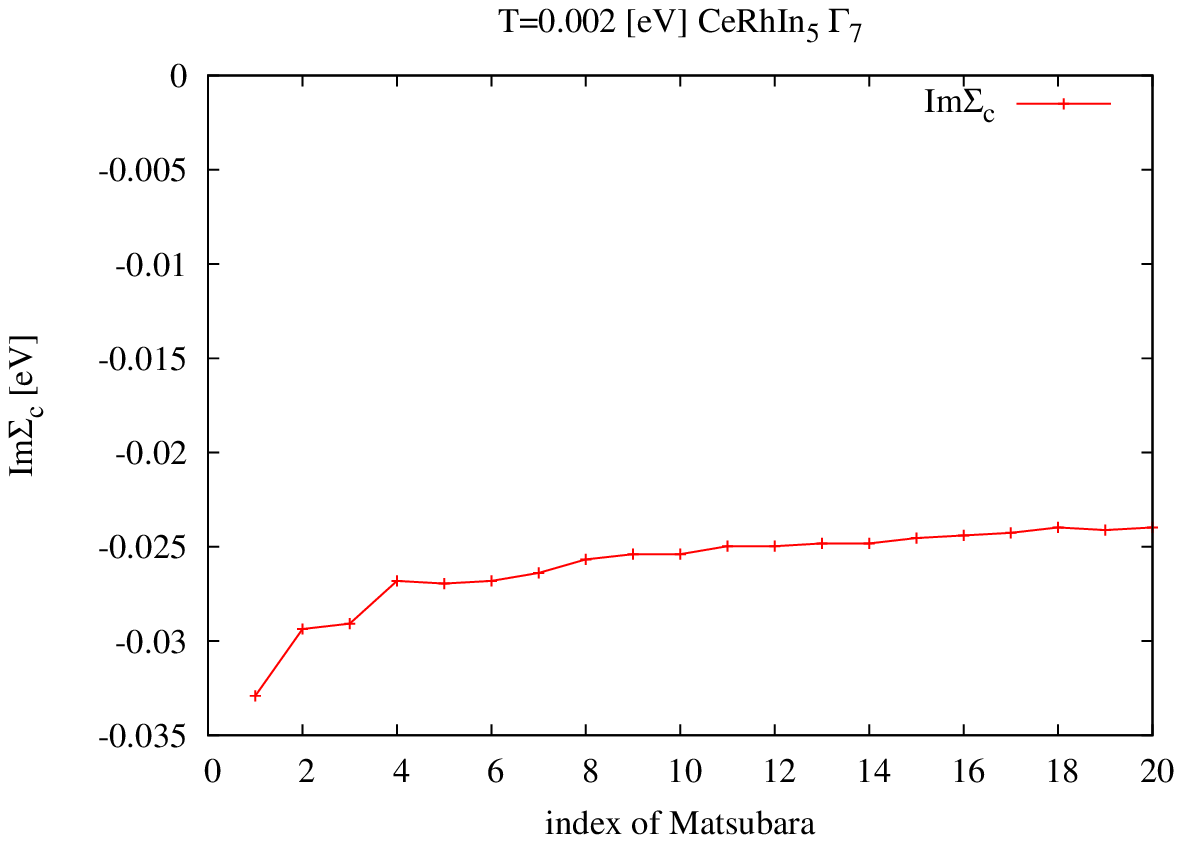}
\caption{(color online) Conduction electron self--energies $\mathfrak{R}%
\Sigma _{c}(\protect\epsilon )\ $(top plot) and $\Im \Sigma _{c}(\protect%
\epsilon )$ (bottom plot) of $\Gamma _{7}$ states for CeRhIn$_{5}$ at
temperature $T=0.002$ eV($\approx $23K$)$ calcultated using the LDA+DMFT
method with the Kondo lattice.}
\label{Sigma_c}
\end{figure}

A general LDA+DMFT\ calculation for this material has been done in the
former work \cite{MatsumotoPRL} with $\epsilon _{f}=-2.5\text{eV}$ and $U=5%
\text{eV}$ which are the typical values for Ce--based compounds. Here we
provide an analysis of its low energy physical properties. Fig. \ref{Sigma_c}
shows calculated conduction electron self--energies,$\mathfrak{\ R}\Sigma
_{c}(\epsilon )\ $and $\Im \Sigma _{c}(\epsilon ),$ for the $\Gamma _{7}$
state from our LDA+DMFT simulation with the Kondo lattice. We clearly see
that the imaginary part of $\Sigma _{c}$ tends to diverge at least to $%
T\approx 23\text{K although }$this behavior could result from the
temperature being not low enough in our simulation. However, a conventional
low--frequency expansion of the self--energy and the connection between $%
z_{f}$ and $z_{c},$ Eq. (\ref{zf2zc}), cannot be utilized to estimate the
quasiparticle residue for the f electrons and the Sommerfeld coefficient $%
\gamma $.

Here we use our mapping method to extract $z_{f}$ exactly as we illustrated
for our model calculation. In this way, we first estimate the quasiparticle
residue $z_{f}$ and, second, evaluate the renormalized density of states at
the Fermi level $\ N(0)_{\text{eff}}=N(0)_{\text{LDA}}/z_{f}$. Then the
Sommerfeld coefficient can be found

\begin{equation}
\gamma =\frac{1}{3}\pi {N_{\text{eff}}(0).}
\end{equation}%
All calculated properties are summarized in Table \ref{CeRhIn5}. It can be
seen that our estimate for $\gamma $ is very close to the experimental value
which indicates that our simulation is sufficiently accurate to describe
this material.

\section{CONCLUSION}

We have studied a mapping of periodic Anderson model to the Kondo lattice
model in the limit of $U\rightarrow \infty $ for single--particle functions
such as the self--energy. The crossover occurs at the values of interaction $%
U=10D$ where the models become equivalent. This allowed us to map the
quasiparticle residue $z_{f}$ of the f--electrons and extract its values
directly from the Kondo lattice model. We applied the method to realistic
heavy fermion system CeRhIn$_{5}$ where our estimates for the Sommerfeld
coefficient agree well with the experiment.

\section*{ACKNOWLEDGMENTS}

We gratefully acknowledge useful discussions with X. G. Wan and Shu--Ting
Pi. The work was supported by DOE NEUP under Contract No. 00088708.

\end{document}